# Sensitivity and Robustness of Quantum Spin-½ Rings to Parameter Uncertainty

Sean O'Neil, *Member, IEEE,* Edmond Jonckheere, *Life Fellow, IEEE,* Sophie Schirmer, *Member, IEEE,* and Frank Langbein, *Member, IEEE*

*Abstract*— Selective transfer of information between spin-1/2 particles arranged in a ring is achieved by optimizing the transfer fidelity over a readout time window via shaping, externally applied, static bias fields. Such static control fields have properties that clash with the expectations of classical control theory. Previous work has shown that there are cases in which the logarithmic differential sensitivity of the transfer fidelity to uncertainty in coupling strength or spillage of the bias field to adjacent spins is minimized by controllers that produce the best fidelity. Here we expand upon these examples and examine cases of both classical and non-classical behavior of logarithmic sensitivity to parameter uncertainty and robustness as measured by the $\mu$ function for quantum systems. In particular we examine these properties in an 11-spin ring with a single uncertainty in coupling strength or a single bias spillage.

## I. INTRODUCTION

Information encoded in networks of coupled spins can propagate without mass or change transport. For example, spintronic devices using nuclear or electron spins confined to quantum dots in 2D electron gas (2DEG) controlled by surface electrodes could overcome limitations imposed by mass or charge transport and hold significant promise for on-chip communication [1]. Linear chains and rings can be used as the components of quantum wires and routers. Due to the complex wave-like propagation of excitations in such networks, efficient transfer of information is non-trivial, necessitating effective control. One way this can be achieved is by energy-landscape shaping via time independent controls, such as voltages applied to the gate electrodes, to alter the energy levels of the electrons confined to the quantum dots [2,3].

Such systems are interesting from a control theory perspective as they exhibit unusual robustness properties. As demonstrated in [4] and [5], when examining the sensitivity of the system to uncertainty in spin couplings or leakage of the nominal bias field from the intended spin to adjacent spins or measuring the robustness of the system's performance to these same uncertainties we observe trends that appear to contradict the classical control limitations imposed by the identity $S + T = I$ where $S$ is the sensitivity transfer matrix and $T$ is the complementary sensitivity transfer matrix. To be more precise, we observe cases in which the probability of successful transfer is maximal while the logarithmic sensitivity is nearly zero, in contradiction to the classical intuition [4].

Additionally, extending the analysis to larger, non-differential uncertainties through μ-analysis reveals instances of anti-classical behavior with the most optimal controllers also being the most robust in many cases [5]. In this paper, we aim to expand upon the results detailed in [5] by examining a larger data set and looking at cases of both classical as well as anti-classical behavior.

## II. BACKGROUND

### A. *Problem Formulation and Structure*

As discussed in detail in [2], we consider a system composed of N spin-$\frac{1}{2}$ particles arranged in a ring with one excitation present between the N spins. We aim to find a control $D$ that maximizes the probability of transfer of the single excitation from a particular spin $n$ to a specific target spin $m$ at a given time $t_f$ or over a time window $[t_f - \delta t, t_f + \delta t]$. We can then identify the state of the quantum system with the excitation localized at the initial spin as the state $|IN\rangle$ and identify the desired final state with the excitation localized at spin $m$ as $|OUT\rangle$. Taking the control $D$ as a diagonal $N \times N$ matrix describing the bias applied to each spin to affect the desired transfer, we see that the system is governed by the equation $|\dot\psi\rangle = -i(H + D)|\psi\rangle$ where $H$ is the single excitation subspace Hamiltonian of the ring (here a constant circulant matrix). The probability of transfer at a given time $t_f$ is then equal to the squared fidelity

$$|\langle OUT|\psi(t_f)\rangle|^2 = |\langle OUT|e^{-i(H+D)t_f}|IN\rangle|^2 = prob. \quad (1)$$

Departing from this nominal model, we consider two categories of perturbations as described in [5]. The first is an uncertainly in the assumed uniform coupling strengths between spins. We model this perturbation as an element $\delta_{k,k+1} S_{k,k+1}$ appended to the nominal Hamiltonian. Here $S_{k,k+1}$ is an $N \times N$ matrix that provides a specific structure to the perturbation with the only non-zero elements being 1's in the $(k, k+1)$ and $(k+1, k)$ positions for $k < N$ and in the $(N, 1)$ and $(1, N)$ positons for $k = N$. Additionally $\delta_{k,k+1}$ provides the size of the perturbation to the nominal coupling strength. The other category of perturbation we consider is that of a leakage of the bias field intended for spin $k$ to its neighbors. We model this perturbation as a term $\delta_k D_k S_k$ added to the Hamiltonian. As before $\delta_k$ provides the size of the perturbation with respect to spillage at spin $k$, and $S_k$ is a

Sean O'Neil is with the Electrical Engineering and Computer Science Department of the United States Military Academy, West Point, NY 10996, USA. (phone: 253-548-5825; e-mail: sean.o'neil@usma.edu).

Dr. Edmond Jonckheere is with the Department of Electrical Engineering at the University of Southern California, Los Angeles, CA 90089, USA. (e-mail: jonckhee@usc.edu).

Dr. Sophie Schirmer is with the College of Science, Swansea University, Swansea Wales, UK. (e-mail: lw1660@gmail.com).

Dr. Frank Langbein is with the School of Computer Science & Informatics, Cardiff University, Cardiff, Wales, UK. (e-mail: LangbeinFC@cf.ac.uk). FCL and SGS are supported by Sêr Cymru NRN AEM grant 82.

matrix that carries the structure of the perturbation. Here $S_k$ is a matrix of all zeros save for $-1$ in the $(k,k)$ position and $\frac{1}{2}$ in the $(k-1,k-1)$ and $(k+1,k+1)$ positions. $D_k$ is a scalar that represents the bias intended for the *k*th spin (i.e. the $(k,k)$ element of the diagonal matrix of controls $D$).

*B. Sensitivity Analysis*

In classical multivariable control, we see the tension between tracking error and logarithmic sensitivity to parameter variation in the identity $S + T = I$. This tension is evident through identification of the log-sensitivity with $T$ through the relation $S^{-1}(dS) = (dL)L^{-1}T$ [4].

To relate this to the quantum system of interest, we first define a "tracking error" in the sense of the difference between the achieved probability of transfer and unity (perfect state transfer). Matters are complicated by the fact that in the case of maximum fidelity we have $|\langle OUT|\psi(t_f)\rangle|^2 = 1$ which implies that $|\psi(t_f)\rangle = e^{i\varphi(t_f)}|OUT\rangle$ for some global phase factor $\varphi(t)$ that is hidden in the computation of the fidelity squared. Therefore, if we take the tracking error as the "closeness" of the state achieved at time $t_f$ to the desired state $|OUT\rangle$, we get from [4]

$$\left\| |OUT\rangle - e^{-i\varphi(t_f)}|\psi(t_f)\rangle \right\|^2 = 2\big(1 - |\langle OUT|\psi(t_f)\rangle|\big) = 2\,err^2 \tag{2}$$

Note that to minimize this error does not require that $|\psi(t_f)\rangle$ approach $|OUT\rangle$ in the sense of an ordinary signal, but that the norm in (2) be minimized with respect to $\varphi(t_f)$, and as such is referred to as the projective tracking error [4].

With this relation of the error to the "reference signal" $|OUT\rangle$, we then see that computing the logarithmic sensitivity of the system to parameter uncertainty is tantamount to taking the derivative $\frac{\partial(err)}{\partial \delta_k}\left(\frac{1}{err}\right)$ which, with $prob = (1 - err)$ allows for concordance between the size of $\left|\frac{\partial(err)}{\partial \delta_k}\left(\frac{1}{err}\right)\right|$ and $\left|\frac{\partial(prob)}{\partial \delta_k}\left(\frac{1}{err}\right)\right| = \left|\frac{\partial(prob)}{\partial \delta_k}\left(\frac{1}{1-(prob)}\right)\right|$.

Now we can evaluate $\frac{\partial(prob)}{\partial \delta_k}$ for a single perturbation $\delta_k$ as per [2]:

$$\frac{\partial |\langle OUT|e^{-i(\tilde H)t_f}|IN\rangle|^2}{\partial \delta_k}$$
$$= -2t_f \sum_{m,n} \langle OUT|\tilde\Pi_m S_k \tilde\Pi_n|IN\rangle \mathrm{sinc}\left(\frac{1}{2}t_f\tilde\omega_{mn}\right)$$
$$\times \sum_l \langle IN|\tilde\Pi_l|OUT\rangle \sin\left(\frac{1}{2}t_f(\tilde\omega_{nl} + \tilde\omega_{ml})\right) \tag{3}$$

Here $\tilde\Pi_l$ and $\tilde\omega_{nl} = \tilde\lambda_n - \tilde\lambda_l$ are taken from the eigendecomposition of $\tilde H = \sum_l \tilde\lambda_l \tilde\Pi_l$ where $\tilde H$ is the perturbed Hamiltonian defined as $H + D + \sum_{k=1}^{N-1} \delta_{k,k+1} S_{k,k+1} + \delta_{N,1} S_{N,1}$ for the case of coupling uncertainty or $H + D + \sum_{k=1}^{N} \delta_k D_k S_k$ for bias spillage. An integral of (3) over the readout window centered on $t_f$ then yields a measure of the windowed fidelity's error to differential parameter variations and thence to the log-sensitivity.

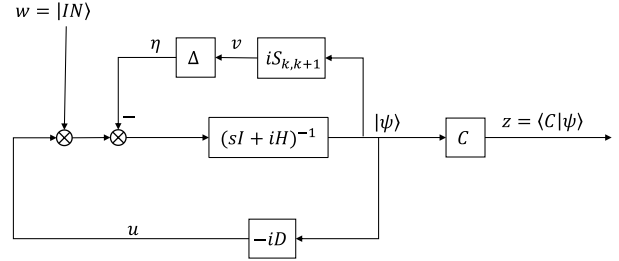

Fig. 1. Block diagram representation of the system with a single uncertainty in the coupling strengths between spins modeled as an inverse additive uncertainty around the plant.

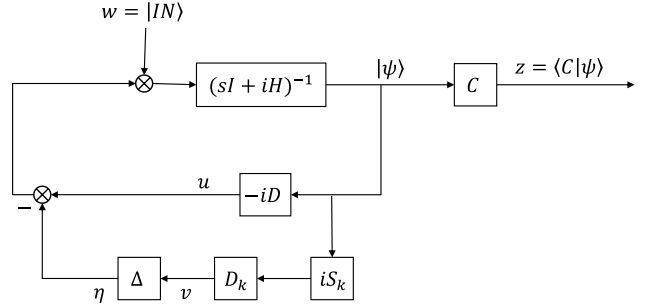

Fig. 2. Block diagram representation of the system with a single uncertainty in the applied bias modeled as an additive uncertainty around the control.

*C. μ-Analysis*

To analyze the robustness of the system via the $\mu$-function we represent the systems described by the two forms of perturbations as in Figs. 1 and 2 in accordance with [6, Chapter 8] where the perturbations are limited to that of a single $\delta$.

*Importantly, note that although we have the controller $-iD$ located in the feedback path, there is no measurement performed on the wavefunction to be compared against a reference signal in order to drive the dynamics.* Rather, the controller alters the energy landscape of the system to modify the natural evolution of the system in a pre-determined manner [2].

Here, we define the initial condition as a disturbance at the plant input so that $w = |IN\rangle$ is our generalized disturbance. We assign the generalized error as in [5] as $z = \langle C|\psi\rangle$ where $C$ is an $(N-1) \times N$ matrix with rows that form a basis for the orthogonal complement of our desired output $|OUT\rangle$. In both cases $\Delta$ is an $N \times N$ diagonal matrix that consists of $\delta_{k,k+1}$ or $\delta_k$ times the identity matrix. $\eta$ and $v$ are signals used to close the loop around the uncertainty that's been "pulled out" of the system. Solving in terms of the generalized inputs and outputs yields the following for the coupling uncertainty and bias spillage cases respectively:

$$\begin{pmatrix} v \\ z \\ \psi \end{pmatrix} = \begin{pmatrix} -iS_{k,k+1}\Phi & iS_{k,k+1}\Phi & iS_{k,k+1}\Phi \\ -C\Phi & C\Phi & C\Phi \\ -\Phi & \Phi & \Phi \end{pmatrix} \begin{pmatrix} \eta \\ w \\ u \end{pmatrix} = P^1 \begin{pmatrix} \eta \\ w \\ u \end{pmatrix} \tag{4}$$

$$\begin{pmatrix} v \\ z \\ \psi \end{pmatrix} = \begin{pmatrix} -iD_k S_k \Phi & iD_k S_k \Phi & iD_k S_k \Phi \\ -C\Phi & C\Phi & C\Phi \\ -\Phi & \Phi & \Phi \end{pmatrix} \begin{pmatrix} \eta \\ w \\ u \end{pmatrix} = P^2 \begin{pmatrix} \eta \\ w \\ u \end{pmatrix} \tag{5}$$

where we use $\Phi = (sI + iH)^{-1}$ to simplify the notation as in [4]. From this point we perform a lower linear fractional transformation as per [6, Chap. 8] to pull the controller into the generalized plant yielding $M^k = F_l(P^k, -iD)$ for $k = 1,2$ and $P^k$ is partitioned appropriately to yield:

$$M_{11}^k = P_{11} - iP_{13}D(I + iP_{33}D)^{-1}P_{31}$$
$$M_{12}^k = P_{12} - iP_{13}D(I + iP_{33}D)^{-1}P_{32}$$
$$M_{21}^k = P_{21} - iP_{23}D(I + iP_{33}D)^{-1}P_{31}$$
$$M_{22}^k = P_{22} - iP_{23}D(I + iP_{33}D)^{-1}P_{32}$$

Finally, we can absorb the structured uncertainty into the plant-controller system yielding $T_{zw} = F_u(M^k, \Delta) = M_{22}^k + M_{21}^k \Delta(I - M_{11}^k \Delta) M_{12}^k$. Closing the loop from $z$ to $w$ with a fictitious, full uncertainty matrix $\Delta_p$ with dimensions consistent with $z$ and $w$ we obtain the system in Fig. 3 where $\tilde{\Delta} = \begin{pmatrix} \Delta & 0 \\ 0 & \Delta_p \end{pmatrix}$ and has an obvious block diagonal structure.

This use of a full uncertainty matrix $\Delta_p$ arises from a constraint of MATLAB's *mussv* function which requires a full uncertainty matrix for calculation of $\mu$ when the perturbations are complex. As $\Delta_p$ closes the loop from the generalized error $z$ to the generalized disturbance $w = |IN\rangle = |\psi(0)\rangle$ we should expect $\Delta_p$ to be structured to only permit influences from an error in the initial state preparation to affect the generalized error, or should be rather sparse with the only non-zero columns corresponding to entries in the state vectors that carry the complex uncertainty in the initial state preparation. As such, the results produced by *mussv* may be overly conservative.

From here we can examine the robust performance of the system in seeking a bound $\beta$ such that $\|T_{zw}\| < \beta$ for all $\|\tilde{\Delta}\| \leq \frac{1}{\beta}$ which from [7, Chap. 10] amounts to finding a lower bound on the function $\mu_{\tilde{\Delta}}(M)$ allowing us to leverage the tools of $\mu$-analysis to determine a measure of robustness of the system. Before proceeding, however, we must state the caveat as in [5] that though classically, nominal and robust stability are prerequisites of robust performance, in this study, the system is not asymptotically stable in the usual sense, as the control is state selective and time-sensitive. So while we presently use the tools of $\mu$-analysis to study robustness of the excitation transfer over a finite time window, it must be kept in mind that other tools may be necessary to study robustness in such a non-classical system generally.

### III. RESULTS

#### A. Simulation Procedure

As in [5] we use the model of an 11-ring with nominal XX-coupling as our system of interest. For this ring size, we consider first the controllers optimized to maximize the transfer fidelity over a window $[t_f - 0.1, t_f + 0.1]$ [2]. For each transfer from $|1\rangle \to |1\rangle$ through $|1\rangle \to |6\rangle$ the previously executed optimization algorithm produces a data set of up to 2000 diagonal controllers $D$ along with the time-averaged probability [8]. With each of these variables ordered by decreasing value of probability, we then use the simulation to test the trend between both log-sensitivity and robustness and the probability of transfer.

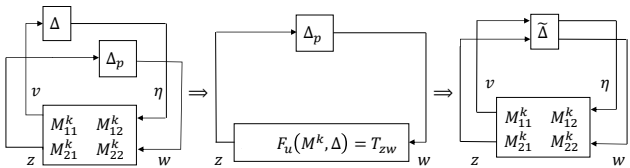

Fig. 3. Transformation of system to allow for $\mu$-analysis with structured perturbation a block-diagonal $\tilde{\Delta}$.

The log-sensitivity is calculated in accordance with II-B for each $D(n)$ optimized for the six possible transfers within the 11-ring, taking into account the 11 possible cases of coupling uncertainty and separate 11 cases of bias spillage for each possible transfer. This produces a total of 132 test cases to measure the relationship between probability of transfer and log-sensitivity.

For the calculation of $\mu_{\tilde{\Delta}}(M(D(n)))$ we begin with each set of 2000 controllers for each of the six possible transfers and use the system set-up detailed in II-C while leveraging MATLAB's *mussv* function to evaluate the lower bound on each $\mu_{\tilde{\Delta}}(M(D(n)))$. As in [5] we evaluate $\Phi = (sI + iH)^{-1}$ at $s = 0$ to reflect that with the input as a constant in time, it is part of the exponential regime $e^{0 \cdot t}$ and thus the output attributable to the input is also part of this regime. Finally, we take the matrix $\tilde{\Delta}$ as an element of $\mathbb{C}^{22 \times 21}$ structured block-diagonally with the upper-left block consisting of $\delta I_{11 \times 11}$ for the model uncertainty and the lower-right block composed of a full $11 \times 10$ matrix. This process is repeated for each test case described above to permit a comparison of robustness and probability.

In addition to the data set described above, and to allow for a continuation of the results detailed in [5] we also consider the set of 1000 controllers optimized to provide maximum fidelity within a shortest time $t_f$ for a $|1\rangle \to |3\rangle$ transfer. These controllers are reordered in descending rank based on their time-averaged probability of transfer via a numerical integration over the period $[t_f - 0.1, t_f + 0.1]$. Then for each case of coupling uncertainty from 1-2 through 11-1 and bias spillage over all eleven spins we compute the log-sensitivity and structured singular value for each of the set of 1000 reordered controllers at our disposal. This provides another 44 test cases for our study though all are limited to a $|1\rangle \to |3\rangle$ transfer for this data set.

#### B. Hypothesis Test and Statistical Analysis

As in [4] and [5], the data gathered in the study is extremely noisy making empirical calculation of trends nearly impossible. Thus, with the large number of test cases at our disposal, we turn to statistical analysis to determine the trend or lack thereof between the metrics of interest: probability vs. log-sensitivity and probability vs. robustness.

We establish the hypothesis test using the Kendall $\tau$ to measure the level of concordance between metrics. We set the null hypothesis $H_0$ to align with the mean of $\tau = 0$, indicating no rank correlation between probability and log-sensitivity or robustness. We take the alternative hypothesis $H_1$ as negative correlation between the same metrics. Thus failure to reject $H_1$ indicates results inconsistent with the expectations of classical control theory.

To provide bounds on the hypothesis test we first note that the sample size in each test case is either 2000 or 1000. With such sample sizes, the Kendall $\tau$ tends to a standard normal distribution [9] under the null hypothesis with a Z-score of $Z_\tau = \frac{\tau}{\sigma_\tau}$ where $\sigma_\tau = \sqrt{\frac{2(2\ell+5)}{9\ell(\ell-1)}}$. Here we use $\ell$ to denote the number of samples (controllers) within the given data set. As such we can set the value of Type I error as $\alpha = 0.05$ in a single-tailed, negative-tailed test using the value of $Z_\tau$ as the test statistic. We then take $Z_\tau < -1.645$ as the indication for rejection of the null hypothesis and a strong indication of non-classical behavior since $\Phi(Z_\tau) = p < 0.05$ under this condition. Furthermore, for each case in which we reject $H_0$ we can associate a power to the test based on associating the true population Kendall $\tau$ with the observed sample Kendall $\tau$, in which case $Z_\tau < -2.486$ would indicate a power of 0.80 or greater for the case of our 2000 controller data sets and $Z_\tau < -2.487$ would provide the same for the 1000 controller set.

We apply this hypothesis to the trends of log-sensitivity versus probability and $\mu$ versus probability for each of the 308 test cases described above, allowing for a decision on the rejection or failure to reject non-classical behavior for each transfer and each type of perturbation based on the p-value calculated above. To get a better result for the overall trends, however, we look to combine the data in such a manner as to indicate the overall decision on the existence of non-classical behavior for the entire set of possible perturbations within each excitation transfer. As such we use Stouffer's method to combine the 11 values of $Z_\tau$ for the coupling uncertainty and bias spillage test cases using $Z_S = \frac{\sum_{k=1}^{11} Z_{\tau k}}{\sqrt{11}}$, allowing for calculation of the overall p-value for each distinct transfer based on a Stouffer p-value of $p_S = \Phi(Z_S)$ [10 and 11].

As a precondition for using Stouffer's method to synthesize p-values, however, it's necessary that each experiment (test case) be independent. We justify the independence among all test cases by the randomness of the numerical optimization scheme [4].

*C. Coupling Uncertainty Results*

The results of the hypothesis test applied to the log-sensitivity and robustness to coupling uncertainty are summarized in Table I. Note that the relationship between probability and both log-sensitivity and $\mu$ reject the null hypothesis with an overall p-value of zero to four decimal places, indicating a very strong negative correlation among the metrics for the transfers $|1\rangle \to |1\rangle$, $|1\rangle \to |2\rangle$, and $|1\rangle \to |3\rangle$. The transfers to the remaining spins then show highly classical behavior in response to the set of coupling uncertainties with p-values of unity to four decimals places. Thus we see that using Stouffer's Method to allow for the combination of p-values almost produces a "zero-one" hypothesis test for the existence of anti-classical behavior with a sharp change in the system behavior as we move from the $|1\rangle \to |3\rangle$ to $|1\rangle \to |4\rangle$ transfers. This agrees with the results of [4] in which we see the highest levels on non-classical behavior to coupling uncertainty in the transfers that are in physical proximity to the initial spin. As the target spin is moved to the antipodal point on the ring, however, we regain the classical relations between probability and log-sensitivity and robustness that one would expect. In Fig. 4 we see an illustration of these non-classical trends as borne out by the statistical tests. In like manner, the relation in Fig. 5 between the probability and log-sensitivity shows that the ring exhibits almost zero sensitivity to parameter variations for those controllers that allow for nearly perfect fidelity, again in contradiction to classical expectations.

On the other hand, in Fig. 6 we see an illustration of the classical behavior for coupling uncertainty between spins 5 and 6 in a $|1\rangle \to |6\rangle$ transfer. Here the hypothesis test rejects the possibility of non-classical behavior with p-values of near unity for both log-sensitivity and $\mu$ versus probability. This classical trend is easily observed from the graph.

Table I: Results of hypothesis test applied to the case of coupling uncertainty. Shaded boxes indicate rejection of the null hypothesis and strong non-classical behavior.

| Coupling Uncertainty Summary - 11 Ring dt-Data | | | | | |
|---|---|---|---|---|---|
| Transf | Probability and $\mu$ | | | Probability and Log Sensitivity | | |
| | Mean $\tau$ | Mean Z | Stouffer p | Mean $\tau$ | Mean Z | Stouffer p |
| 1->1 | -0.3084 | -20.6694 | 0.0000 | -0.3481 | -23.3296 | 0.0000 |
| 1->2 | -0.0436 | -2.9201 | 0.0000 | -0.4411 | -29.5646 | 0.0000 |
| 1->3 | -0.0683 | -4.5793 | 0.0000 | -0.1643 | -11.0142 | 0.0000 |
| 1->4 | 0.1670 | 11.1923 | 1.0000 | 0.2407 | 16.1319 | 1.0000 |
| 1->5 | 0.2258 | 15.1345 | 1.0000 | 0.6637 | 44.4840 | 1.0000 |
| 1->6 | 0.2229 | 14.9395 | 1.0000 | 0.7058 | 47.3063 | 1.0000 |

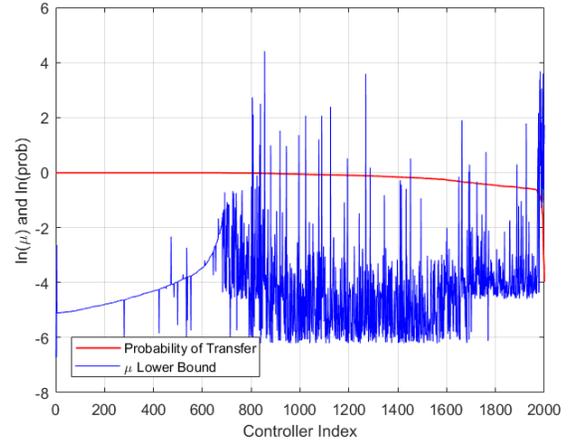

Fig. 4. Plot of the logarithm of $\mu$ versus logarithm of probability for a 1 to 2 transfer with coupling uncertainty between spins 11 and 1 illustrating the overall trend of decreasing robustness with decreasing probability.

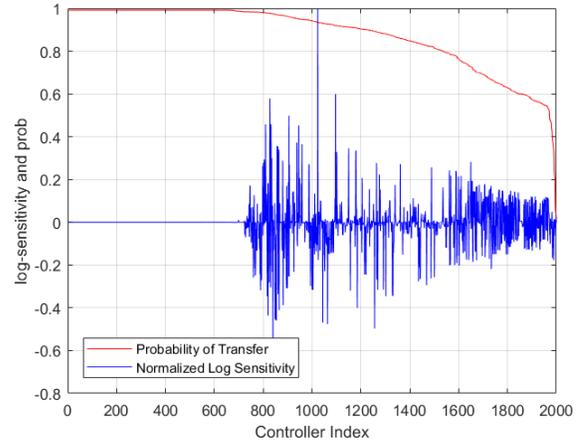

Fig. 5. Plot of the log-sensitivity versus probability for coupling uncertainty between spins 11 and 1 in an excitation transfer from spin 1 to 2 illustrating the overall negative trend between the two metrics. Note that the controllers that allow for almost perfect fidelity also have vanishing sensitivity, in contradiction to the expectations of classical control.

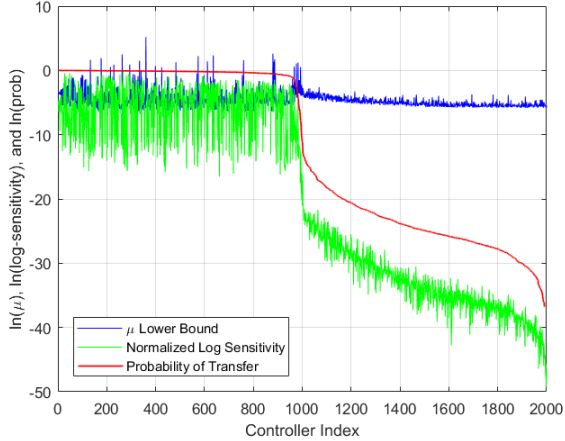

Fig. 6. Consolidated plot of metrics for the case of a 1 to 6 transfer and coupling uncertainty between spins 5 and 6. Note the very close concordance between the log-sensitivity and the probability, especially in the region for controllers between 1000 and 2000.

For the case of the 1000 controller data set used in [5], Table II provides the results of the hypothesis test applied to each case of coupling uncertainty for the $|1\rangle \rightarrow |3\rangle$ transfer. For the 22 available test cases, we see that 20 present non-classical trends and of these 20, 18 cases reject the null-hypothesis with a power of 0.80 or greater, indicating very strong non-classical behavior. We do note, however, that the only test cases that fail to reject the null hypothesis with this power threshold are those with coupling uncertainty or bias spillage on the spins in the shortest physical path between the initial and target spin, indicating a trend toward more classical behavior with perturbations in these locations.

*D. Bias Spillage Results*

As with the coupling uncertainty results, the results of the hypothesis test when taken over bias spillage are summarized in Table III. Here we see rejection of the null hypothesis in only three situations: between both $\mu$ and log-sensitivity and probability for the case of localization about the initial spin and between log-sensitivity and probability for the case of a $|1\rangle \rightarrow |2\rangle$ transfer. Though the overall results of the hypothesis test indicate far more classical behavior for perturbations in the form of bias spillage, we do again see that the excitation transfers with a target spin closest to the initial spin exhibit the most non-classical behavior.

As Fig. 7 reveals for the case of localization of the excitation about $|1\rangle$, both $\mu$ and log-sensitivity steadily increase as the probability of transfer decreases. This should be somewhat expected as these cases are indicative of Anderson localization, perhaps the most non-classical behavior possible in a quantum ring.

Finally, as an illustration of the classical behavior indicated by acceptance of the null-hypothesis, we can refer to the graph of Fig. 8. Here it is clear that both $\mu$ and log-sensitivity decrease in concordance with the probability.

Table II: Results of hypothesis test for the case of a 1 to 3 transfer with the 1000-controler data set and taken across all possibilities of a single coupling uncertainty. Note that the only test cases that fail to reject the null hypothesis with a power of at least 0.80 are those in which the coupling uncertainty is physically located between the initial and target spins. (Here $\mu$ Actual denotes the true population mean of the metric.)

| Coupling Uncertainty | $\tau$ for Prob and $\mu$ | Z-Score | p | Accept or Reject Classical Limitations | Would Power > 0.80 Under $\tau = \mu$ Actual |
|---|---|---|---|---|---|
| 1-2 | -0.0506 | -2.394 | 0.0083 | Reject | No |
| 2-3 | -0.04576 | -2.166 | 0.0151 | Reject | No |
| 3-4 | -0.1137 | -5.383 | 0 | Reject | Yes |
| 4-5 | -0.1385 | -6.558 | 0 | Reject | Yes |
| 5-6 | -0.1601 | -7.580 | 0 | Reject | Yes |
| 6-7 | -0.1633 | -7.732 | 0 | Reject | Yes |
| 7-8 | -0.1557 | -7.372 | 0 | Reject | Yes |
| 8-9 | -0.1633 | -7.732 | 0 | Reject | Yes |
| 9-10 | -0.1601 | -7.580 | 0 | Reject | Yes |
| 10-11 | -0.1387 | -6.567 | 0 | Reject | Yes |
| 11-1 | -0.1219 | -5.772 | 0 | Reject | Yes |
| Stouffer Statistics: | -20.1518 | 0.0000 | | Reject | |

| Coupling Uncertainty | $\tau$ for Prob and Log Sensitivity | Z-Score | p | Accept or Reject Classical Limitations | Would Power > 0.80 Under $\tau = \mu$ Actual |
|---|---|---|---|---|---|
| 1-2 | 0.1947 | 9.219 | 1 | Accept | |
| 2-3 | 0.1947 | 9.219 | 1 | Accept | |
| 3-4 | -0.1166 | -5.521 | 0 | Reject | Yes |
| 4-5 | -0.4138 | -19.59 | 0 | Reject | Yes |
| 5-6 | -0.4393 | -20.80 | 0 | Reject | Yes |
| 6-7 | -0.3971 | -18.80 | 0 | Reject | Yes |
| 7-8 | -0.4075 | -19.29 | 0 | Reject | Yes |
| 8-9 | -0.3971 | -18.80 | 0 | Reject | Yes |
| 9-10 | -0.4393 | -20.80 | 0 | Reject | Yes |
| 10-11 | -0.4138 | -19.59 | 0 | Reject | Yes |
| 11-1 | -0.1166 | -5.521 | 0 | Reject | Yes |
| Stouffer Statistics: | -39.2791 | 0.0000 | | Reject | |

Table IV summarizes the results of the hypothesis test applied to the 1000-controller set specific to a $|1\rangle \rightarrow |3\rangle$ transfer. We note the overall mixed results of the hypothesis test in these cases but also again see the trend of more classical behavior as either the bias spillage or coupling uncertainty is in physical proximity to the transfer path.

IV. CONCLUSION

We have demonstrated that in examining the log-sensitivity and robustness of quantum rings controlled by static fields to maximize the probability of transfer of a single excitation that the limits imposed by classical control need not necessarily apply in all cases. In particular we see a general trend of greater non-classicality for transfers between spins in relatively close physical proximity.

Table III: Results of hypothesis test applied to the case of bias spillage. Shaded boxes indicate rejection of the null hypothesis and strong non-classical behavior.

| Bias Spillage Summary - 11 Ring dt-Data | | | | | | |
|---|---|---|---|---|---|---|
| Trans | Probability and $\mu$ | | | Probabillity and Log Sensitivity | | |
| | Mean $\tau$ | Mean Z | Stouffer p | Mean $\tau$ | Mean Z | Stouffer p |
| 1->1 | -0.1151 | -7.7136 | 0.0000 | -0.1720 | -11.5289 | 0.0000 |
| 1->2 | 0.0451 | 3.0204 | 1.0000 | -0.3522 | -23.6038 | 0.0000 |
| 1->3 | 0.0096 | 0.6442 | 0.9837 | 0.1229 | 8.2375 | 1.0000 |
| 1->4 | 0.1662 | 11.1362 | 1.0000 | 0.4160 | 27.8787 | 1.0000 |
| 1->5 | 0.1345 | 9.0153 | 1.0000 | 0.7015 | 47.0175 | 1.0000 |
| 1->6 | 0.1363 | 9.1347 | 1.0000 | 0.7508 | 50.3181 | 1.0000 |

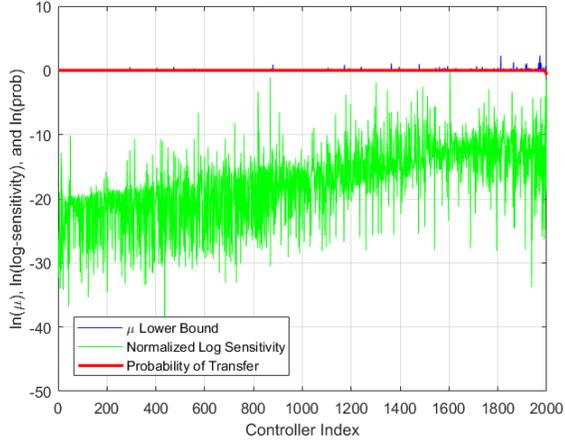

Fig. 7. Consolidated plot of metrics for localization at the initial spin and bias spillage on spin 6. Note the steady increase in the log sensitivity as the probability of transfer decreases from unity and the sharp uptick in $\mu$.

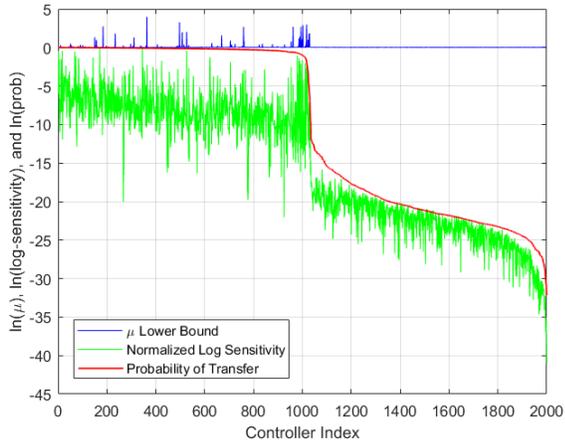

Fig. 8. Consolidated plot of metrics for a transfer from spin 1 to spin 5 and bias spillage on spin 3. Here we see a decrease of both log-sensitivity and $\mu$ in concert with decreasing probability, save for a spike in both metrics around controller index 1000.

We also note that for cases in which the physical location of the uncertainty in coupling strength or bias spillage is in proximity to the excitation transport path, the results more closely follow those anticipated by classical control. Paradoxically, when the source of uncertainty is physically located on the opposite side of the ring from the excitation transfer, we are more likely to see non-classical trends.

Looking forward, it's necessary to extend these results beyond that of an 11-ring to see if these trends can be generalized to systems of arbitrary rings with arbitrary transfers. Finally, it still remains to formulate a model that explains the change from non-classical to classical behavior as the target spins moves to the anti-podal points of the ring.

REFERENCES

[1] D. Loss and D. DiVincenzo. Quantum Computing with Quantum Dots. *Physical Review Letters*, A 57, 120, 1998.

Table IV: Results of hypothesis test for the case of a 1 to 3 transfer with the 1000-controller data set and taken across all possibilities of a single bias spillage. (Here $\mu$ Actual denotes the true population mean of the metric.)

| Spillage Spin | $\tau$ for Prob and $\mu$ | Z-Score | p | Accept or Reject Classical Limitations | Would Power > 0.80 Under $\tau$ = $\mu$ Actual |
|---|---|---|---|---|---|
| 1 | -0.006192 | -0.2931 | 0.3846 | Accept | |
| 2 | -0.03959 | -1.874 | 0.0304 | Reject | |
| 3 | -0.02711 | -1.283 | 0.0996 | Accept | |
| 4 | 0.2018 | 9.555 | 1 | Accept | |
| 5 | -0.2133 | -10.10 | 0 | Reject | Yes |
| 6 | -0.2507 | -11.87 | 0 | Reject | Yes |
| 7 | -0.2430 | -11.50 | 0 | Reject | Yes |
| 8 | -0.2430 | -11.50 | 0 | Reject | Yes |
| 9 | -0.2507 | -11.87 | 0 | Reject | Yes |
| 10 | -0.2134 | -10.10 | 0 | Reject | Yes |
| 11 | 0.2062 | 9.763 | 1 | Accept | |
| Stouffer Statistics: | -15.3988 | 0.0000 | Reject | | |

| Spillage Spin | $\tau$ for Prob and Log Sensitivity | Z-Score | p | Accept or Reject Classical Limitations | Would Power > 0.80 Under $\tau$ = $\mu$ Actual |
|---|---|---|---|---|---|
| 1 | 0.4057 | 19.21 | 1 | Accept | |
| 2 | 0.4107 | 19.44 | 1 | Accept | |
| 3 | 0.4057 | 19.21 | 1 | Accept | |
| 4 | 0.3812 | 18.05 | 1 | Accept | |
| 5 | 0.1541 | 7.296 | 1 | Accept | |
| 6 | -0.06087 | -2.882 | 0.0019 | Reject | Yes |
| 7 | -0.0523 | -2.476 | 0.0066 | Reject | |
| 8 | -0.0523 | -2.476 | 0.0066 | Reject | |
| 9 | -0.0608 | -2.878 | 0.00199 | Reject | Yes |
| 10 | 0.1541 | 7.296 | 1 | Accept | |
| 11 | 0.3812 | 18.05 | 1 | Accept | |

[2] S. G. Schirmer, E. Jonckheere, and F.C. Langbein. Design of Feedback Control Laws for Information Transfer in Spintronic Networks, available at arxiv:1607.05294.

[3] F. C. Langbein, S. G. Schirmer, and E. Jonckheere. Time Optimal Information Transfer in Spintronic Networks. In *IEEE Conference on Decision and Control*, pp. 6454-6459, Osaka, Japan, December 2015.

[4] E. Jonckheere, S. G. Schirmer, and F.C. Langbein. Jonckheere-Terpstra Test for Nonclassical Error Versus Log-Sensitivity Relationship of Quantum Spin Network Controllers, available at arxiv:1612.02784.

[5] E. A. Jonckheere, S. G. Schirmer, and F.C. Langbein. Structured Singular Value Analysis for Spintronics Network Information Transfer Control. *IEEE Transactions on Automatic Control*, in press 2017, available at DOI:10.1109/TAC.2017.2714623, arxiv:1706.03247.

[6] S. Skogestad and I. Postlethwaite. *Multivariable Feedback Control: Analysis and Design*. John Wiley and Sons Ltd, West Sussex, England, 2005.

[7] K. Zhou and J.C. Doyle. *Essentials of Robust Control*. Prentice Hall, Upper Saddle River, NJ, 1998

[8] Frank C Langbein, Sophie G Schirmer, Edmond Jonckheere. Static Bias Controllers for XX Spin-1/2 Rings. Data set, figshare, 2016. DOI:10.6084/m9.figshare.3485240.v1

[9] Herve Abdi, The Kendall Rank Correlation Coefficeint. *Encyclopedia of Measurement and Statistics*. Sage, Thousand Oaks, CA, 2007.

[10] M.C. Whitlock. Combining Probability from Independent Tests: The Weighted Z-Methd is Superior to Fisher's Approach. *Journal of Evolutionary Biology*. 18, pp.1368-1373, 2005.

[11] K. Tsuyuzaki. "Stouffer.test: Stouffer's Weighted Z-Score (Inverse Normal Method)." in metaSeq Version 1.12.0 Available at: https://www.rdocumentation.org/packages/metaSeq/versions/1.12.0/topics/Stouffer.test [accessed August 2017].